\documentclass[submission, Proceedings]{SciPost}

\hypersetup{
    colorlinks,
    linkcolor={red!50!black},
    citecolor={blue!50!black},
    urlcolor={blue!80!black}
}

\usepackage[bitstream-charter]{mathdesign}
\DeclareSymbolFont{usualmathcal}{OMS}{cmsy}{m}{n}
\DeclareSymbolFontAlphabet{\mathcal}{usualmathcal}

\begin{document}

\begin{flushright}
LTH 1263
\end{flushright}

\begin{center}{\Large \textbf{
Complementarity of Lepton-Charge and Forward-Backward Drell-Yan Asymmetries for Precision Electroweak Measurements and Quark Density Determinations\\
}}\end{center}

\begin{center}
J. Fiaschi\textsuperscript{1$\star$},
F. Giuli\textsuperscript{2,3},
F. Hautmann\textsuperscript{4,5} and
S. Moretti\textsuperscript{6}
\end{center}

\begin{center}
{\bf 1} Department of Mathematical Sciences, University of Liverpool, Liverpool L69 3BX, United Kingdom
\\
{\bf 2} CERN, CH 1211 Geneva, Switzerland
\\
{\bf 3} University of Rome Tor Vergata and INFN, Sezione di Roma 2, Via della Ricerca Scientifica 1, 00133 Roma, Italy
\\
{\bf 4} Elementaire Deeltjes Fysica, Universiteit Antwerpen, B 2020 Antwerpen, Belgium
\\
{\bf 5} Theoretical Physics Department, University of Oxford, Oxford OX1 3PU, United Kingdom
\\
{\bf 6} School of Physics and Astronomy, University of Southampton, Highfield, Southampton SO17 1BJ, United Kingdom
\\
* juri.fiaschi@liverpool.ac.uk
\end{center}

\begin{center}
\today
\end{center}


\definecolor{palegray}{gray}{0.95}
\begin{center}
\colorbox{palegray}{
  \begin{tabular}{rr}
  \begin{minipage}{0.1\textwidth}
    \includegraphics[width=22mm]{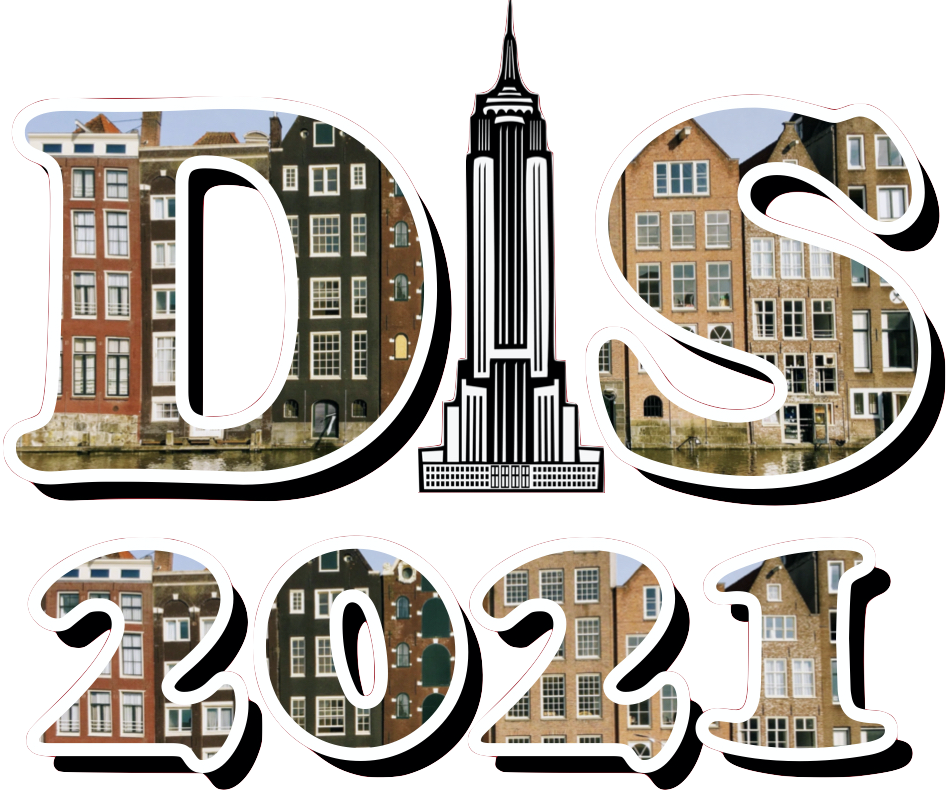}
  \end{minipage}
  &
  \begin{minipage}{0.75\textwidth}
    \begin{center}
    {\it Proceedings for the XXVIII International Workshop\\ on Deep-Inelastic Scattering and
Related Subjects,}\\
    {\it Stony Brook University, New York, USA, 12-16 April 2021} \\
    \doi{10.21468/SciPostPhysProc.?}\\
    \end{center}
  \end{minipage}
\end{tabular}
}
\end{center}

\section*{Abstract}
{\bf
We address the impact of future measurements of charged and neutral current Drell-Yan (DY) asymmetries and their combination on Parton Distribution Functions (PDFs) uncertainties.
We quantify the reduction of PDF uncertainties using the QCD tool xFitter.
We examine the effects of such reduced PDF errors on both Standard Model (SM) and Beyond SM (BSM) observables.
}

\section{Introduction}
\vspace{-1em}
Exploiting the data collected during the ongoing LHC Run-III and future High Luminosity (HL) programmes, not only the sensitivity to BSM physics will sensibly improve, but also very precise determinations of SM observables, such as $\theta_W$ and $m_W$, will be performed, which will be competitive in accuracy with measurements from lepton colliders and Tevatron.

Such statistically precise measurements will be dominated by systematic uncertainties, with one of their largest components coming from non-perturbative effects encoded in the PDFs.
Hence improving our knowledge of the proton structure is crucial for research at current and future colliders.
In the attempt to reduce this source of uncertainty, ATLAS and CMS analysis focused on the determination of Electroweak (EW) parameters at the LHC systematically adopt profiling and reweighting techniques to constrain the PDFs~\cite{Sirunyan:2018swq,ATLAS:2018gqq}.

It has been shown that precise measurements of well understood processes such as DY di-lepton production are very profitable for PDF determinations~\cite{Fu:2020mxl,Deng:2020sol,Willis:2018yln}.
Recent studies have assessed the impact on PDFs from various observables related to the neutral DY process, such as the Forward-Backward Asymmetry ($A_{FB}$)~\cite{Accomando:2017scx,Accomando:2018nig,Abdolmaleki:2019qmq} or the longitudinal polarization coefficient $A_0$~\cite{Amoroso:2020fjw}, further remarking the constraining power of the clean di-lepton final state.

Here we summarise the main results of Ref.~\cite{Fiaschi:2021okg}, where we have studied the constraints on PDFs and their uncertainties from future measurements of the lepton-charge asymmetry ($A_W$) in combination with $A_{FB}$ data, and the consequent improvements in the precision of SM parameters determinations as well as in the sensitivity of specific BSM constructions featuring additional charged and neutral broad resonances in their spectrum.

\section{Complementarity of $A_W$ and $A_{FB}$}
\vspace{-1em}
We make use of the public code {\tt{xFitter}}~\cite{Alekhin:2014irh} to quantify the reduction of PDF uncertainties by profiling~\cite{Paukkunen:2014zia} the Hessian CT14NNLO PDF set~\cite{Hou:2019efy}.
The implementation of the $A_{FB}$ distribution and its fiducial cuts follows the description given in Ref.~\cite{Abdolmaleki:2019qmq}.
The $A_W$ observable as function of the charged lepton pseudorapidity, has been implemented in a similar manner~\cite{Fiaschi:2021okg}, including the fiducial cuts as described in the ATLAS analysis in Ref.~\cite{Aad:2019rou} and QCD NNLO corrections.
Pseudodata for the two observables have been generated assuming two reference integrated luminosities of 300 and 3000 fb$^{-1}$, corresponding to end of Run-III and HL goals respectively.

\begin{figure}[h]
\begin{center}
\includegraphics[width=0.23\textwidth]{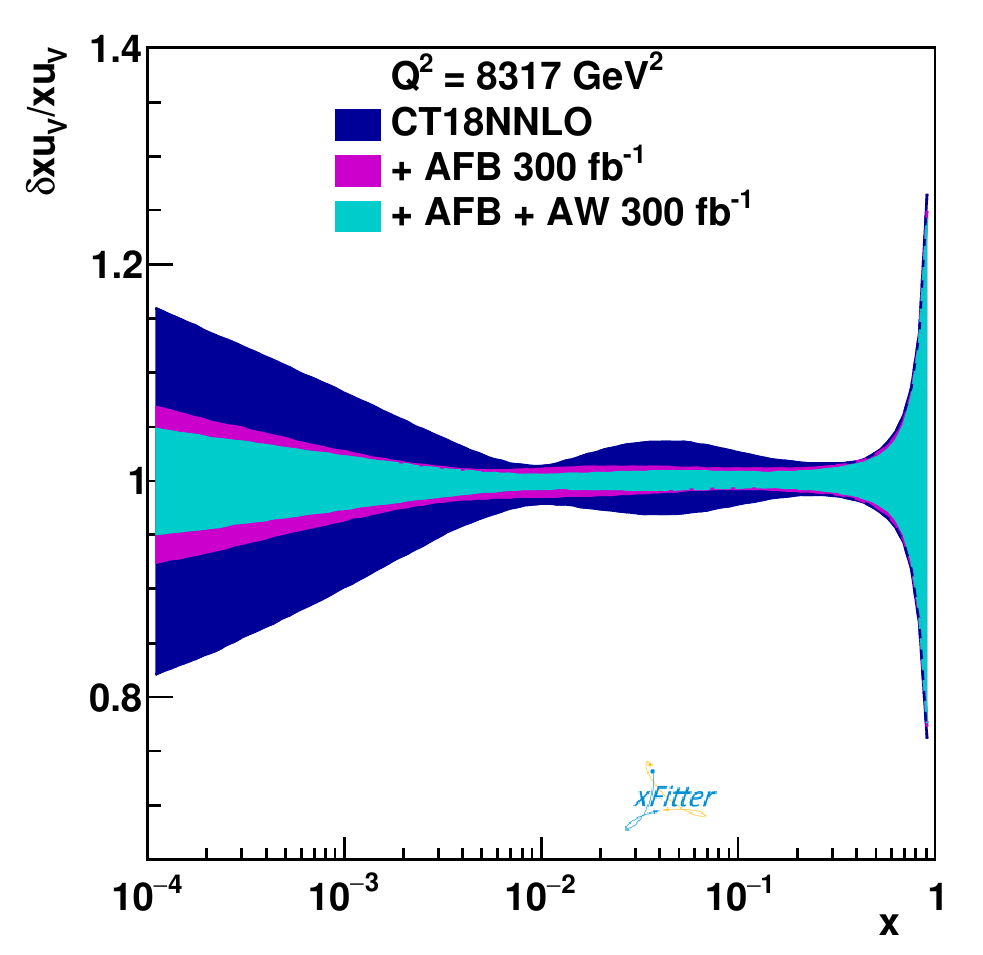}
\includegraphics[width=0.23\textwidth]{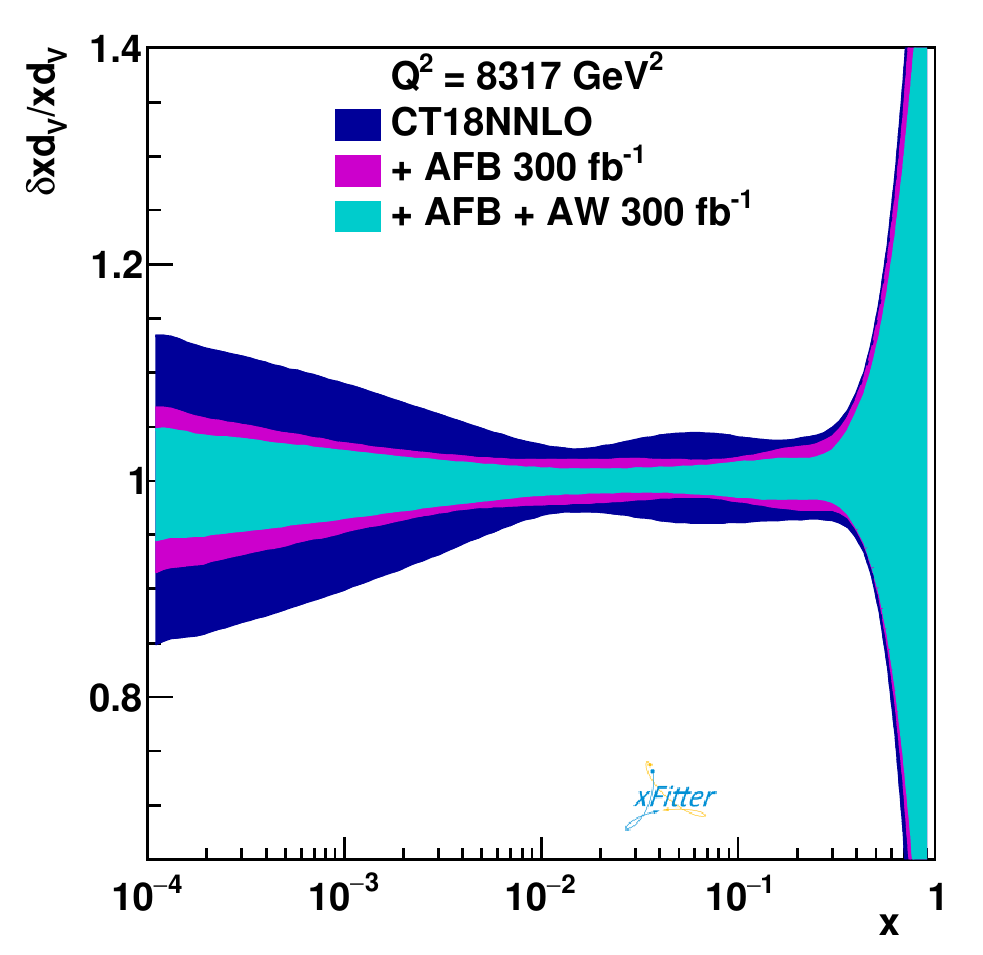}
\includegraphics[width=0.23\textwidth]{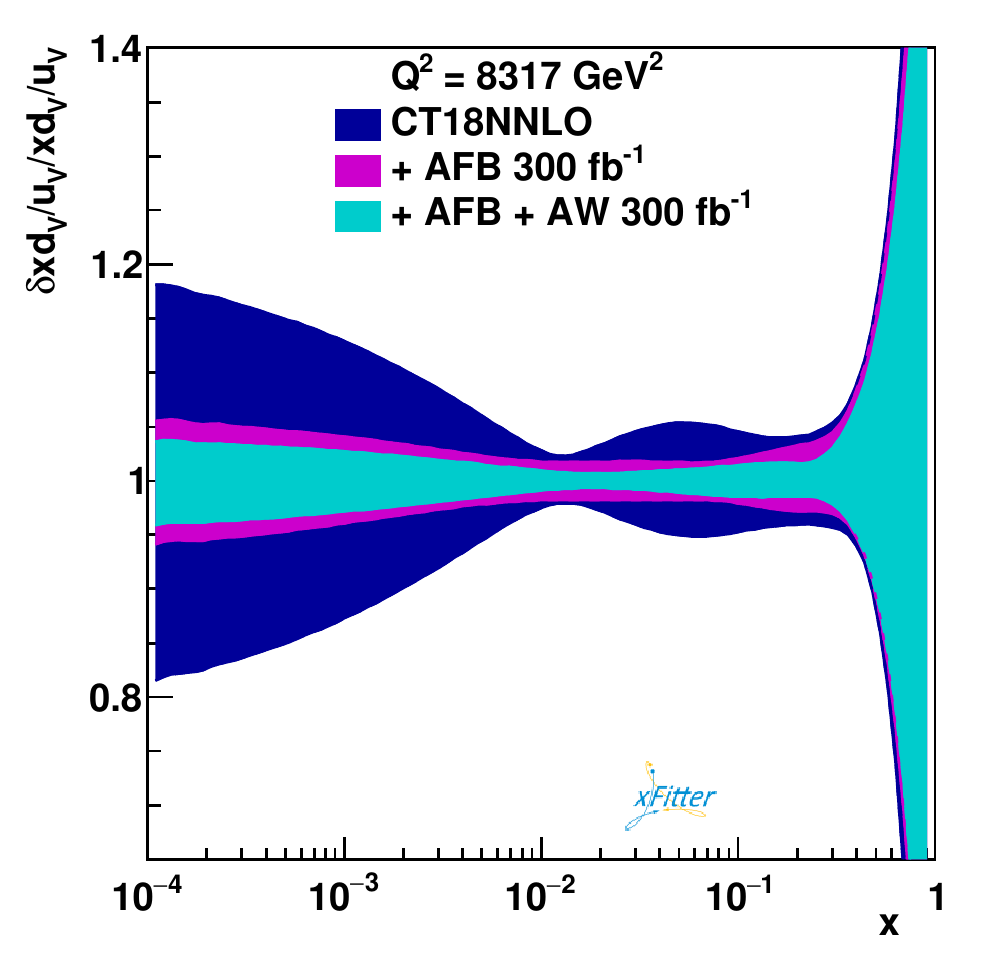}\\
\includegraphics[width=0.23\textwidth]{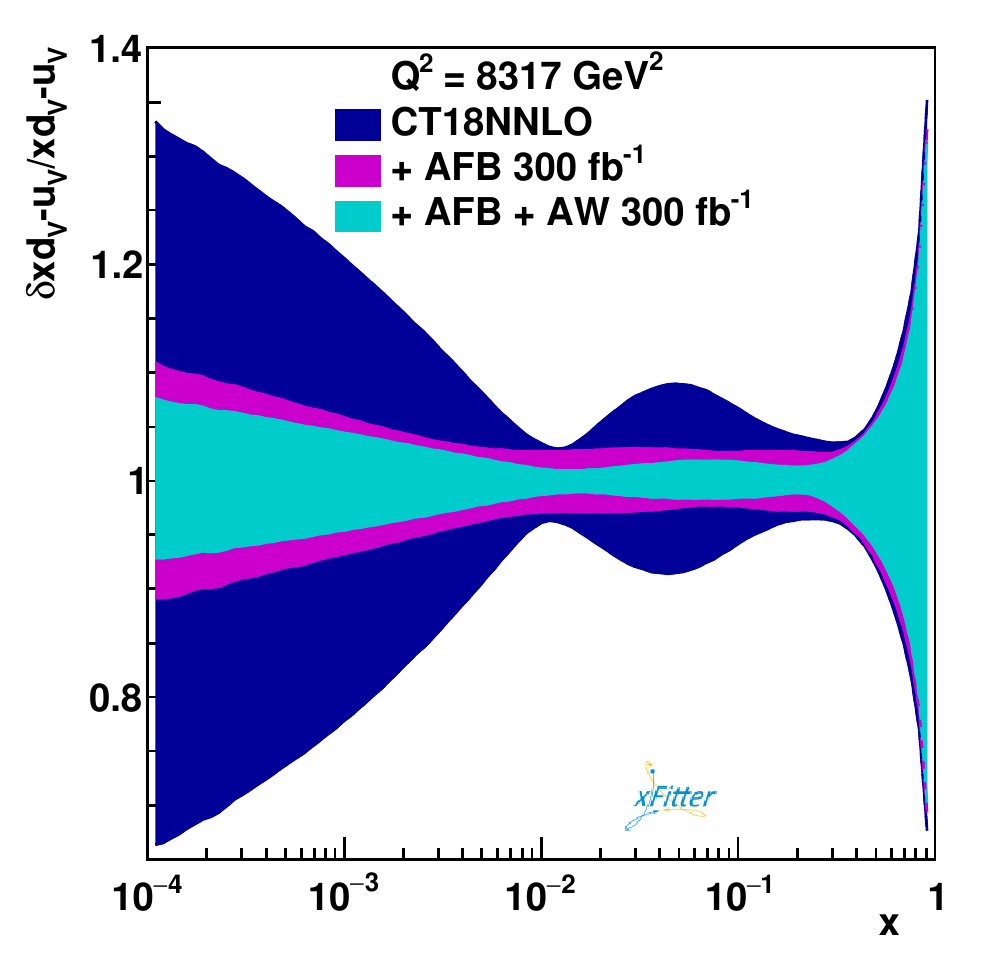}
\includegraphics[width=0.23\textwidth]{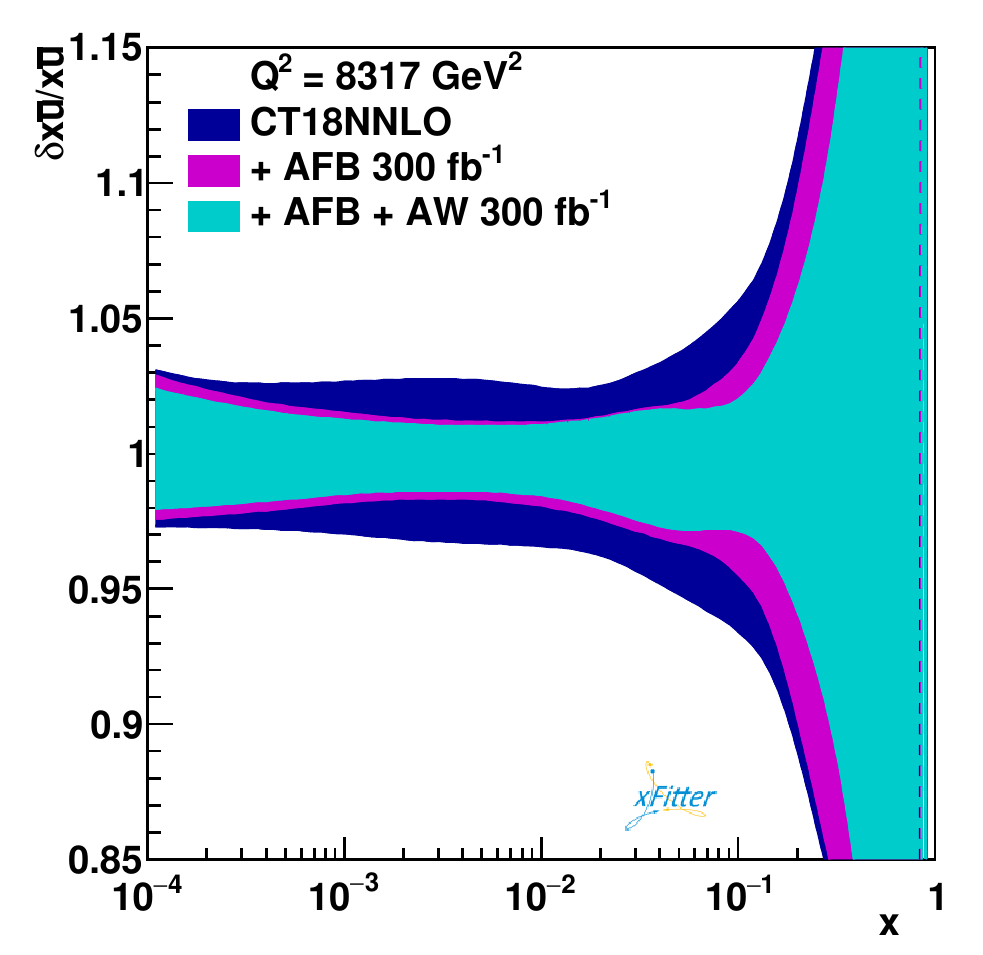}
\includegraphics[width=0.23\textwidth]{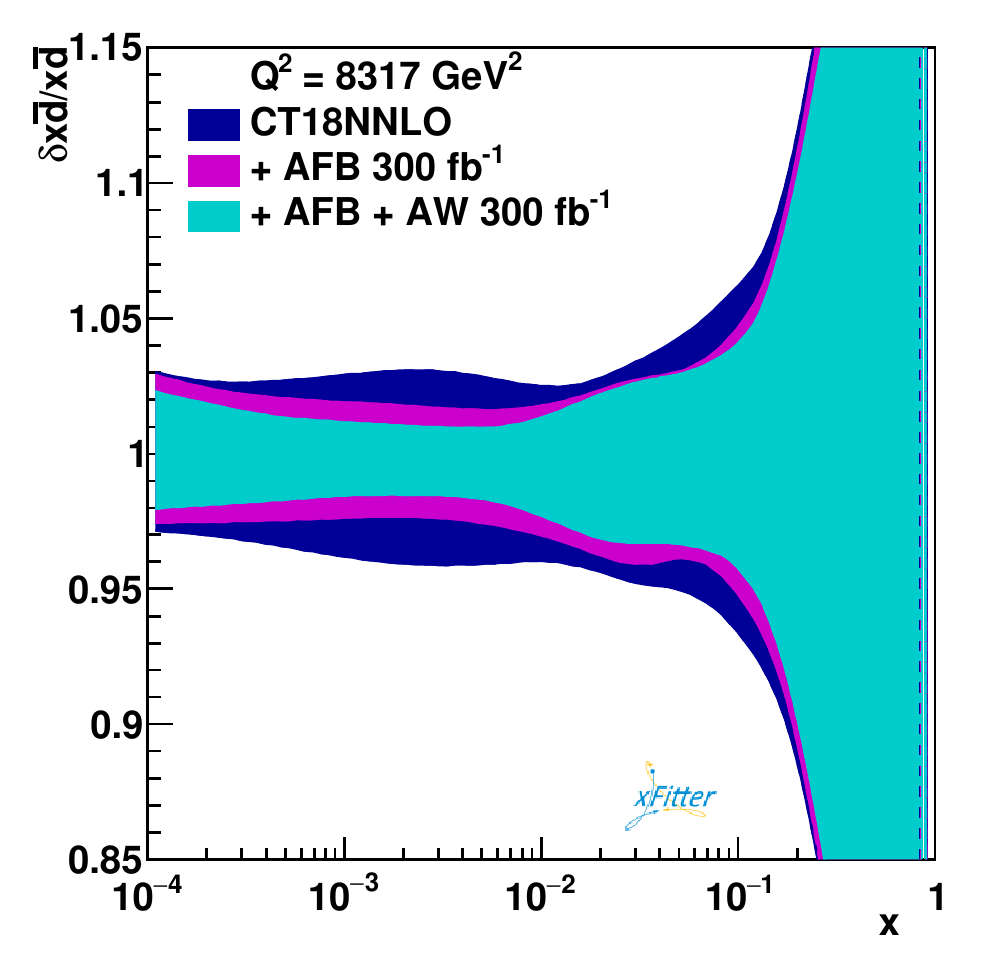}
\end{center}
\caption{Original CT18NNLO~\cite{Hou:2019efy} (blue) and profiled distributions of selected PDFs using either $A_{FB}$ (pink) or both $A_{FB}$ and $A_W$ (cyan) pseudodata corresponding to an integrated luminosity of 300 fb$^{-1}$.}
\label{fig:AW_AFB_comb_300}
\end{figure}

\begin{figure}[h]
\begin{center}
\includegraphics[width=0.23\textwidth]{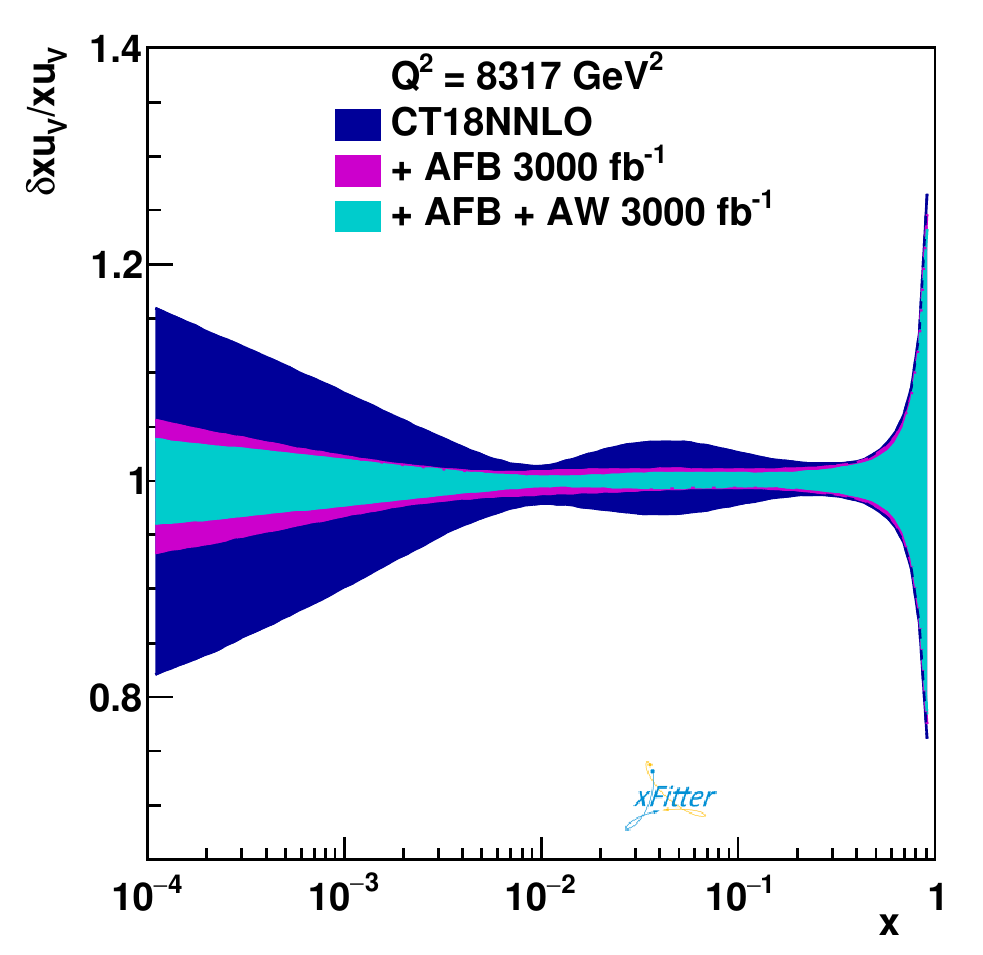}
\includegraphics[width=0.23\textwidth]{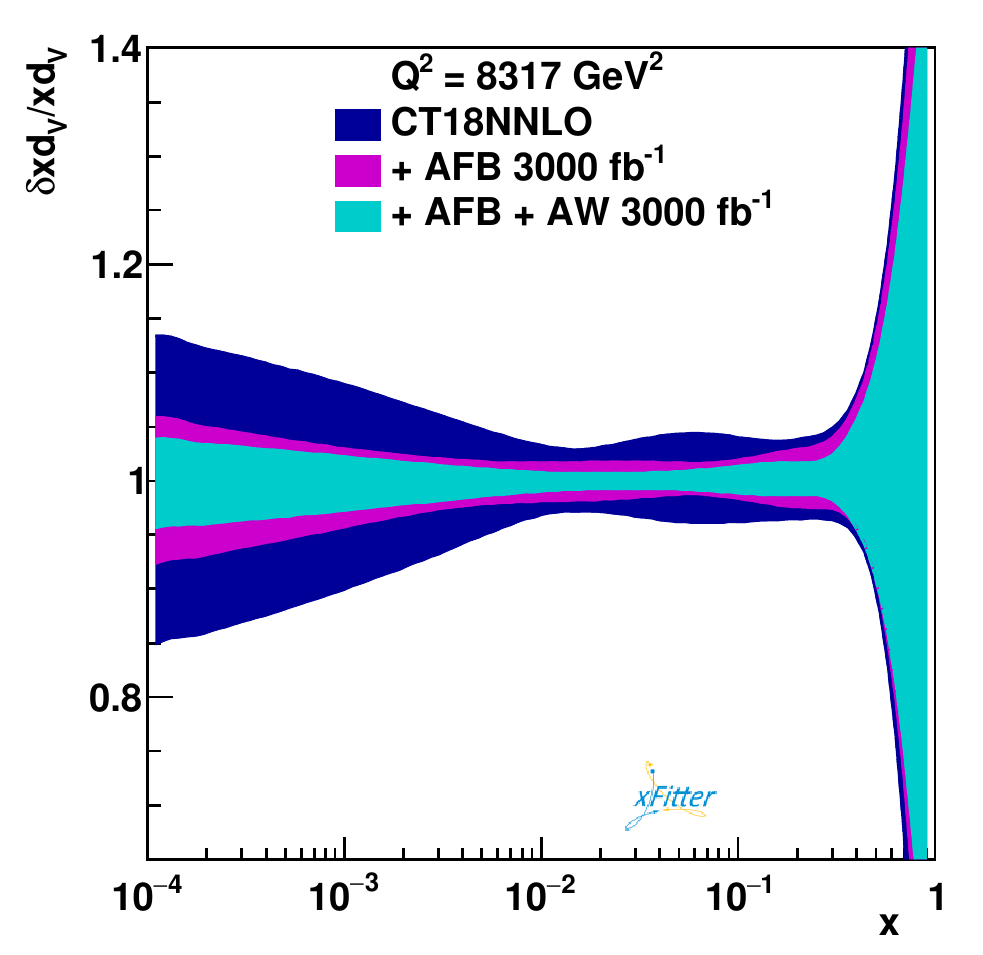}
\includegraphics[width=0.23\textwidth]{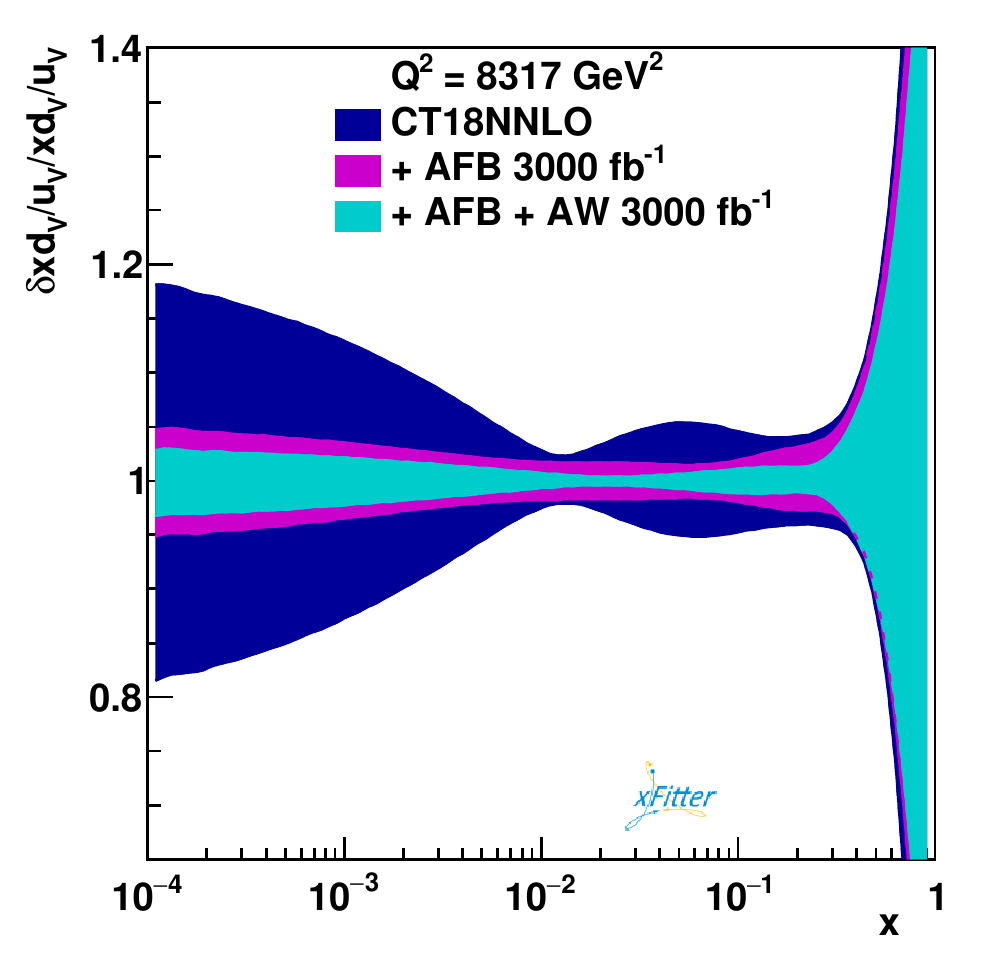}\\
\includegraphics[width=0.23\textwidth]{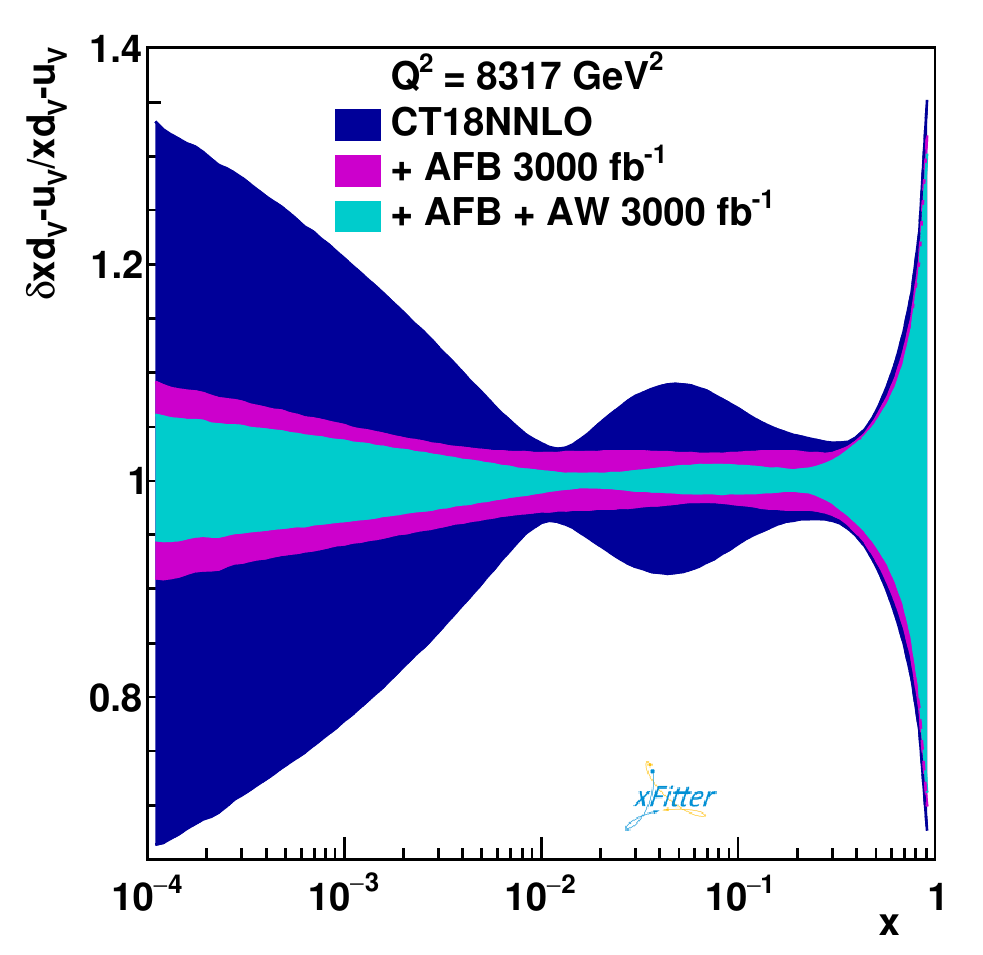}
\includegraphics[width=0.23\textwidth]{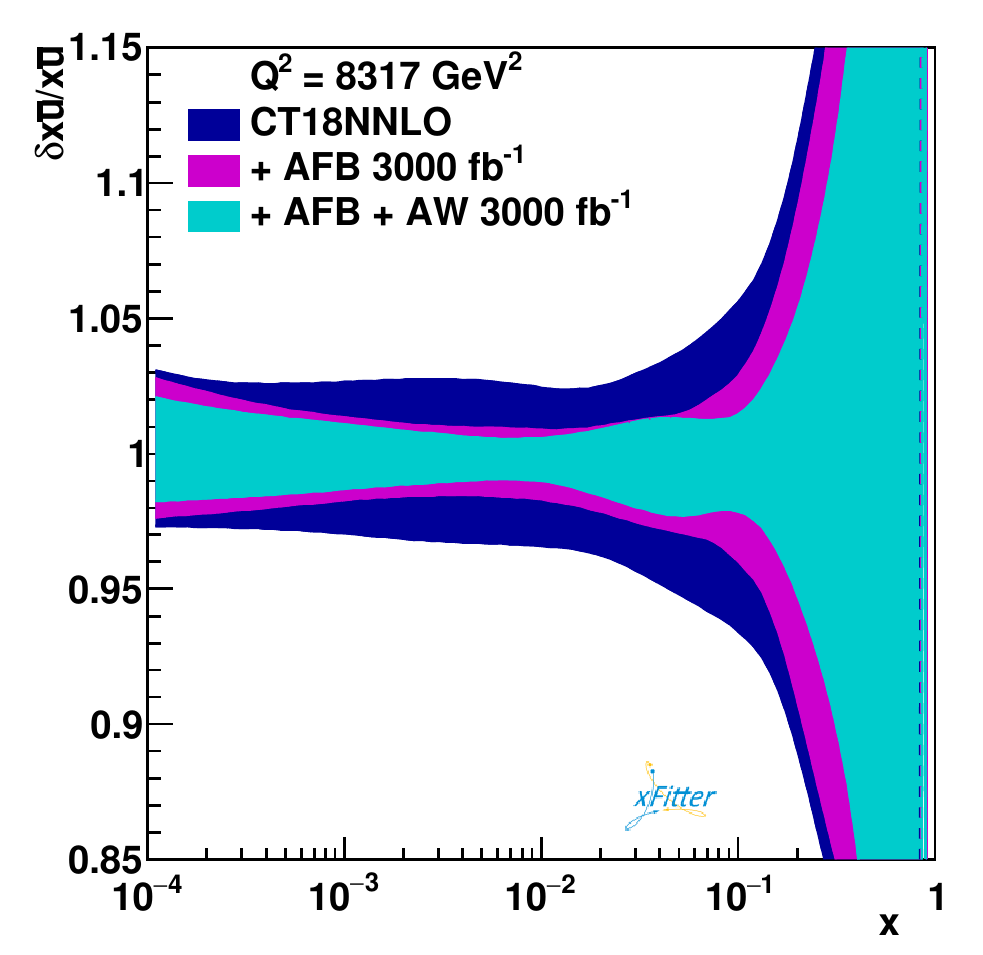}
\includegraphics[width=0.23\textwidth]{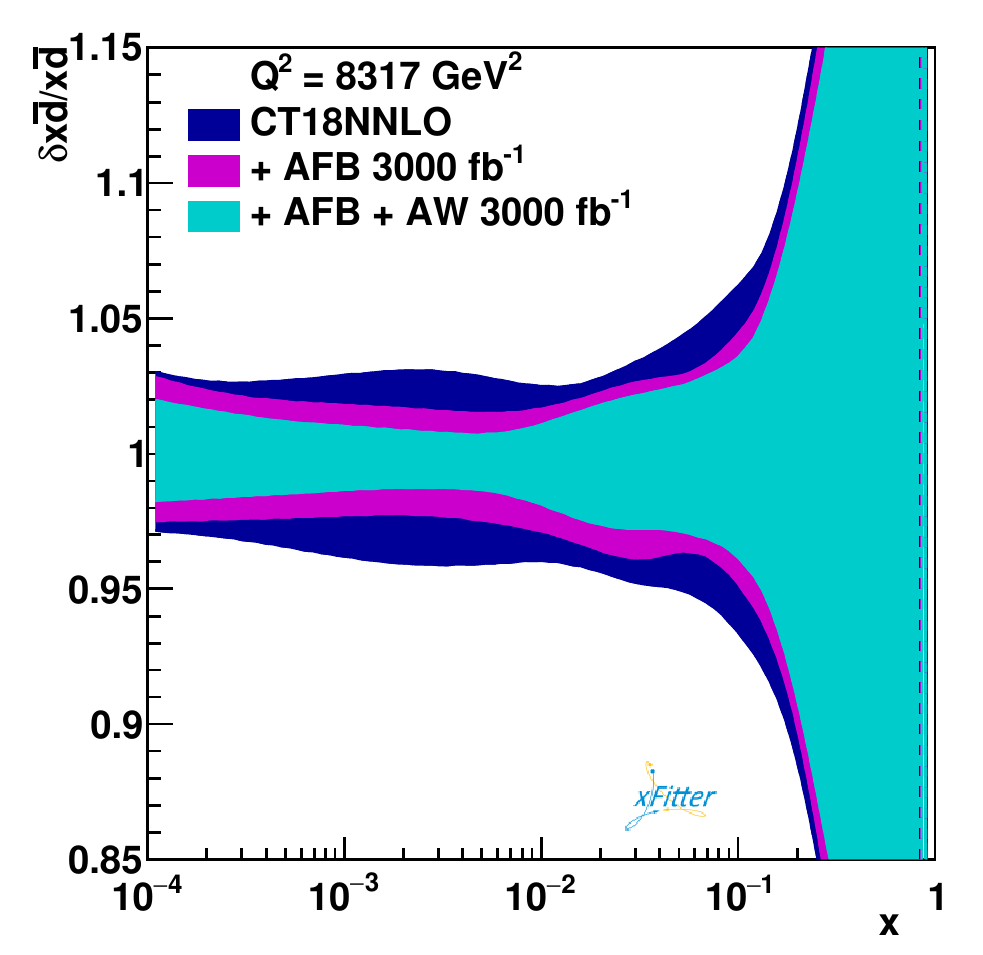}
\end{center}
\caption{Same as in Fig.~\ref{fig:AW_AFB_comb_300} but for an integrated luminosity of 3000 fb$^{-1}$.}
\label{fig:AW_AFB_comb_3000}
\end{figure}

The resulting profiled PDFs are shown in Figs.~\ref{fig:AW_AFB_comb_300} and~\ref{fig:AW_AFB_comb_3000} for the two luminosity scenarios.
The reduction of uncertainties from the inclusion of the two observables pseudodata corresponds to the shrinking of the uncertainty error bands.
This is particularly visible in the valence $u$ and $d$ quark PDFs and even more in their linear combination $d_V - u_V$.
A more moderate contraction of the PDF uncertainty bands is also visible in the antiquark $\bar{u}$ and $\bar{d}$ PDFs.

\section{Impact of PDF uncertainty reduction}
\vspace{-1em}

\subsection{SM measurements}
\begin{figure}[h]
\begin{center}
\includegraphics[width=0.33\textwidth]{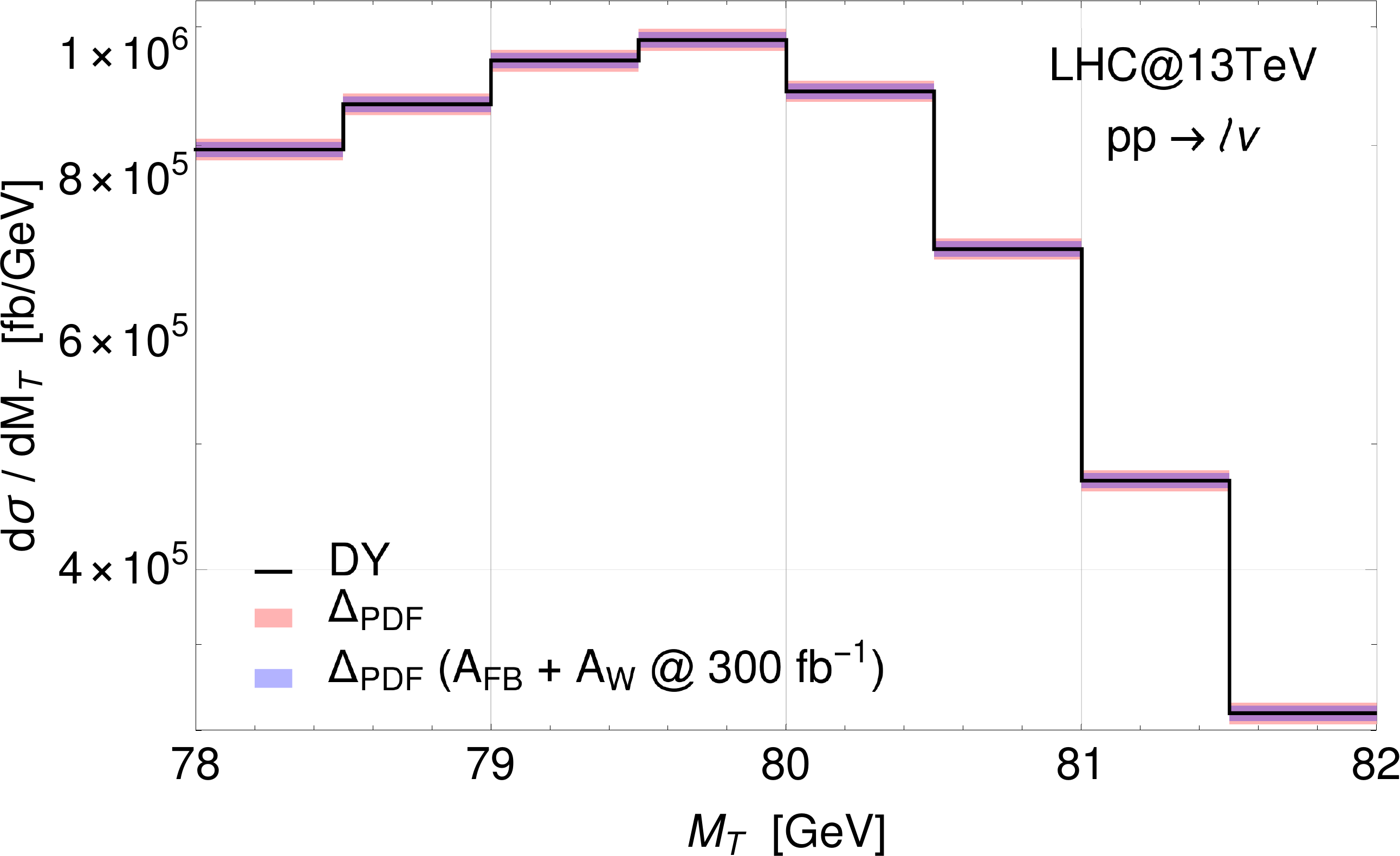}
\includegraphics[width=0.5\textwidth]{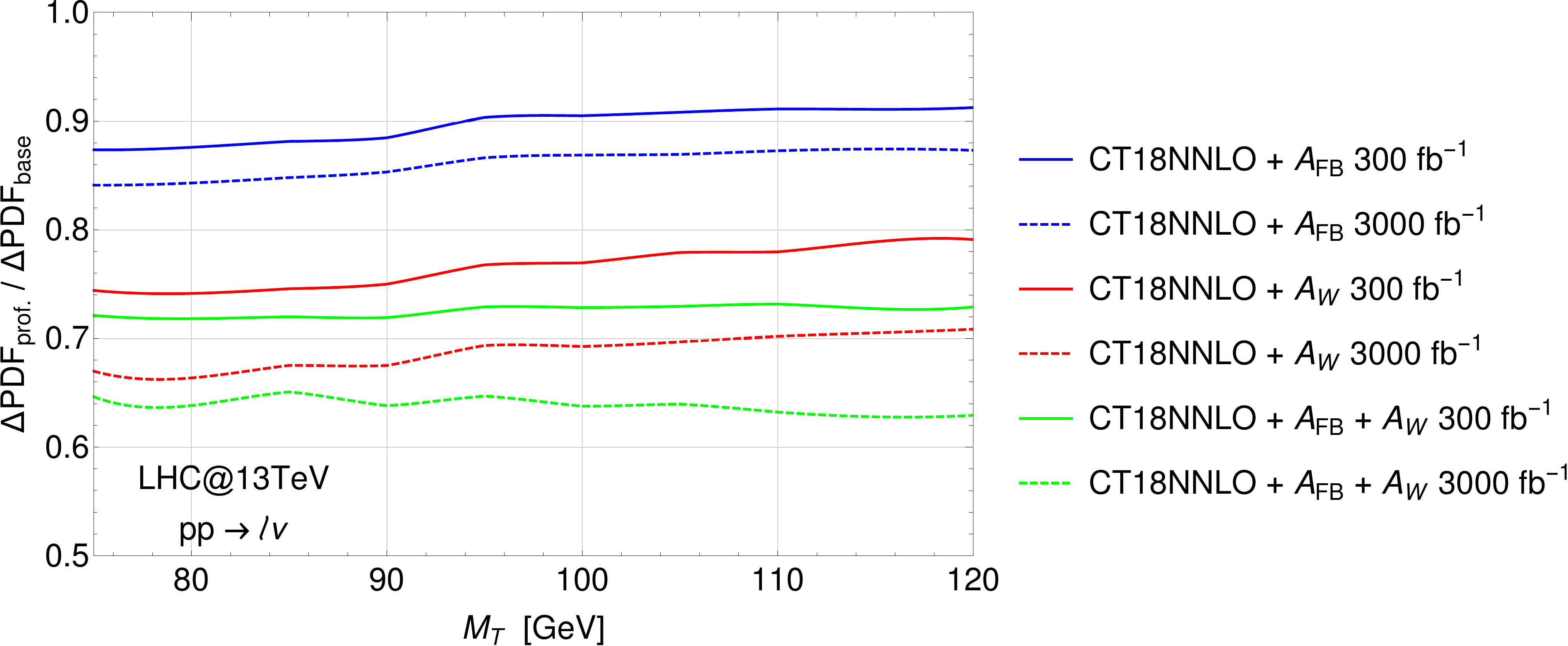}
\end{center}
\caption{(left) Charged-current DY transverse mass distribution with PDF error; (right) relative improvement of PDF error on the transverse mass spectrum due to profiling based on $A_{FB}$, $A_W$ and their combination.}
\label{fig:MT_PDF_error}
\end{figure}

\vspace{-0.5em}
The improvement on PDF determination will have important consequences in the estimation of EW SM parameters.
The measurement of the $W$ boson mass for instance is an important consistency test for the SM, and in this task the LHC has achieved a competitive level of precision with respect to previous experiments.
This quantity is generally extracted from the lepton transverse momentum in the charged DY channel, or from the transverse mass spectrum in the same channel.
The latter is shown on the left plot of Fig.~\ref{fig:MT_PDF_error} together with its statistical and PDF uncertainty after the profiling.
The curves in the plot on the right represent the relative improvement of PDF uncertainty after profiling with respect to the original PDF set using sets of pseudodata with different luminosities of $A_{FB}$, $A_W$ and their combination.
The PDF uncertainty would be reduced by about 12\% (16\%) using $A_{FB}$ pseudodata at 300 (3000) fb$^{-1}$, by about 26\% (43\%) using $A_W$ pseudodata at 300 (3000) fb$^{-1}$, and by about 28\% (46\%) combining the two pseudodata sets at 300 (3000) fb$^{-1}$.

\subsection{BSM searches}
\vspace{-0.5em}
At larger masses (i.e. in the multi-TeV region), the determination of the PDFs is an important factor in BSM physics searches.
Standard analysis for hunting heavy BSM resonances generally assume a Breit-Wigner (Jacobian) peak shape for the $Z^\prime$ ($W^\prime$) signal.
If the resonance is broad, the usual bump hunt procedure is superseded by counting experiments, where an excess of events is sought over an estimated SM background expectation.
In this context, the PDF uncertainty plays an important role in assessing the sensitivity to such BSM signals as its reduction would improve exclusion bounds or could lead to an early discovery~\cite{Accomando:2019ahs}.

\begin{figure}[h]
\begin{center}
\includegraphics[width=0.49\textwidth]{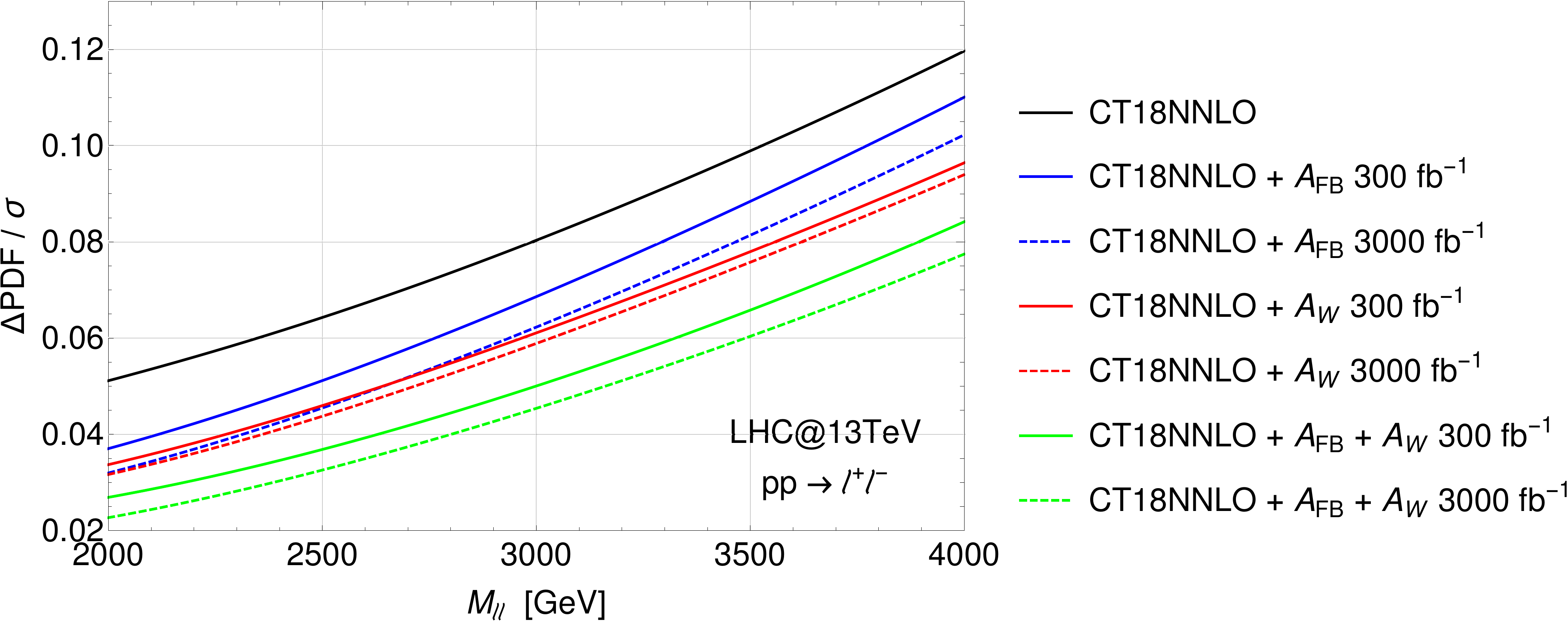}
\includegraphics[width=0.49\textwidth]{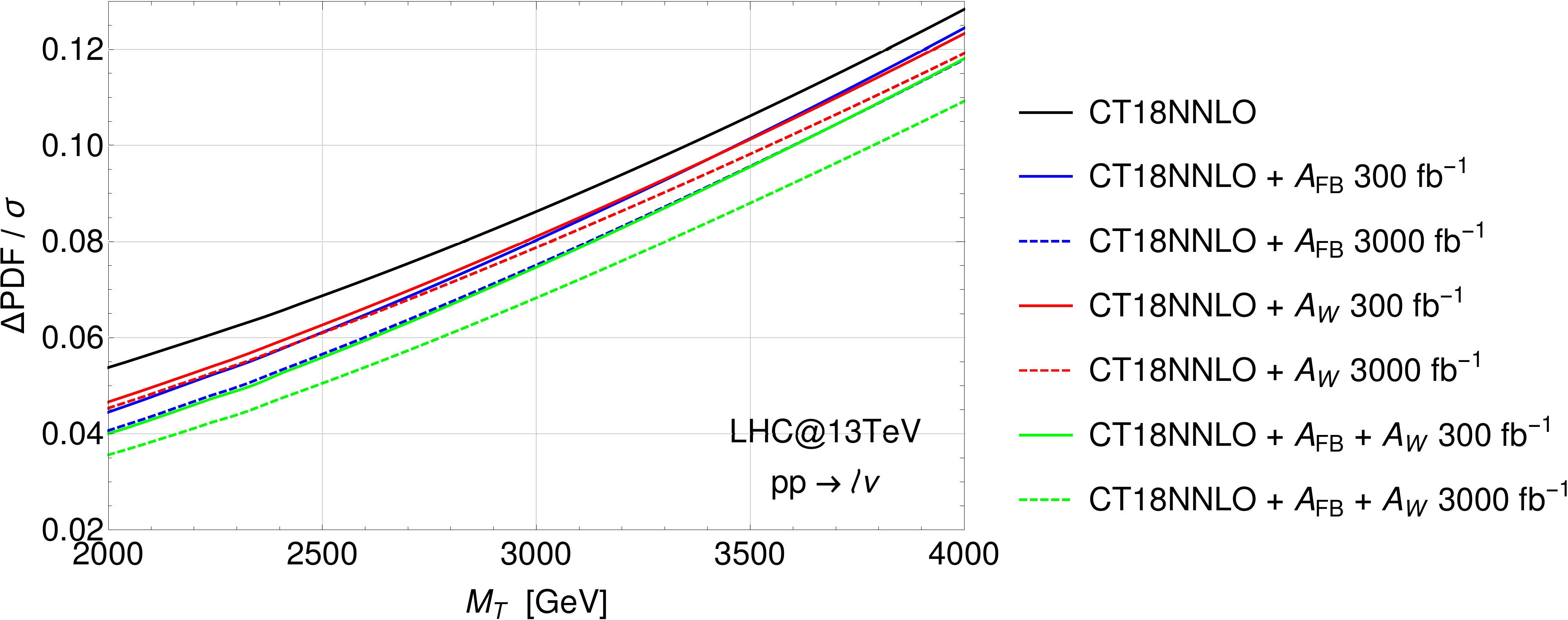}
\end{center}
\caption{Relative PDF error on the invariant mass spectrum of the neutral DY channel (left) and on the transverse mass spectrum of the charged DY channel (right).}
\label{fig:DY_PDF_error}
\end{figure}

Fig.~\ref{fig:DY_PDF_error} shows the relative PDF uncertainty on the cross section of the original and profiled PDF sets using various sets of pseudodata, in the multi-TeV region where broad BSM resonances might still be hiding.
The distributions in the plot on the left are function of the di-lepton invariant mass in the DY neutral channel, while in the plot on the right they are function of the transverse mass in the charged DY channel.

\section{Conclusion}
\vspace{-1em}
DY data in the neutral and charged channels from Run-III and HL stages of the LHC will be an important ingredient in future PDF fits.
Here we have assessed the impact of the combination of the $A_W$ and $A_{FB}$ observables by profiling the CT18NNLO PDF set with pseudodata sets corresponding to 300 and 3000 fb$^{-1}$ integrated luminosity.
We have shown that this data provides a significant reduction of valence quarks PDF uncertainties, and a moderate one in the anti-quark PDFs.
In turn, this improvement can be transferred to future determinations of SM parameters, such as to the extraction of the $W$-boson mass from the charged DY transverse mass spectrum, as well as to searches for BSM heavy bosons decaying into light leptons, where the reduction of PDF uncertainties is particularly beneficial in the case of broad resonances.

\section*{Acknowledgements}
\vspace{-1em}
We thank S.~Amoroso, S.~Camarda and A.~Glazov for many useful discussions. FH acknowledges the hospitality and support of DESY, Hamburg and CERN, Theory Division. We acknowledge the use of the IRIDIS High Performance Computing Facility, and associated support services, at the University of Southampton, in the completion of this work.
\vspace{-1em}
\paragraph{Funding information}
SM is supported in part through the NExT Institute and STFC Consolidated Grant No. ST/L000296/1.
JF is supported by STFC under the Consolidated Grant No. ST/T000988/1.

\bibliography{AFBvsAW_DIS.bib}

\nolinenumbers

\end{document}